\documentclass[12pt,a4paper,dvips]{article}
\usepackage{cite,mcite}
\usepackage{graphicx}
\graphicspath{{/afs/cern.ch/l3/paper/example/figs/}}
\newlength{\capindent}
\newlength{\capwidth}
\newlength{\figwidth}
\newcommand{\icaption}[2][!*!,!]{\hspace*{\capindent}%
  \begin{minipage}{\capwidth}
    \ifthenelse{\equal{#1}{!*!,!}}%
      {\caption{#2}}%
      {\caption[#1]{#2}}
  \end{minipage}}
\def\COST{\mathrm{\rm \cos\theta\;}}
\newcommand{\EE}{\ensuremath{\mathrm{e}^+ \mathrm{e}^-}}
\newcommand{\TeV}{\ensuremath{\mathrm{Te\kern -0.12em V}}}

\newcommand {\Be}{\begin{equation}}
\newcommand {\Ee}{\end{equation}}

\newcommand {\eqref}[1]{equation~(\ref{#1})}

\newcommand {\Figref}[1]{Figure~\ref{fig:#1}}

\renewcommand{\thefootnote}{\fnsymbol{footnote}}

\begin{document}

\begin{titlepage}
\begin{flushright} 
hep-ph/0002172 \\
February 16, 2000
\end{flushright}

\vspace*{3.0cm}

\begin{center} {\Large \bf
       Search for TeV Strings and New Phenomena\\
\vspace*{0.15cm}
       in Bhabha Scattering at LEP2}

\vspace*{2.0cm}
  {\Large
  Dimitri Bourilkov\footnote{\tt e-mail: Dimitri.Bourilkov@cern.ch}
  }

\vspace*{1.0cm}
  Institute for Particle Physics (IPP), ETH Z\"urich, \\
  CH-8093 Z\"urich, Switzerland
\vspace*{4.3cm}
\end{center}

%
%
%
\begin{abstract}
A combined analysis of the data on Bhabha scattering 
at centre-of-mass energies 183 and 189 GeV from the
LEP experiments ALEPH, L3 and OPAL is performed to search
for effects of TeV strings in quantum gravity models with
large extra dimensions. 
No statistically significant deviations from the Standard
Model expectations are observed and lower limit on the 
string scale $\rm M_S = 0.631\ TeV$
at 95~\% confidence level is derived.
The data are used to set lower limits on the scale of
contact interactions ranging from 4.2 to 16.2 TeV depending
on the model.
In a complementary analysis we derive an upper limit
on the electron size of $\rm 2.8\cdot 10^{-19}\ m$
at 95~\% confidence level.

\end{abstract}

\vspace*{1.0cm}

\end{titlepage}

\renewcommand{\thefootnote}{\arabic{footnote}}
\setcounter{footnote}{0}
%
%
\section*{Introduction}

The Standard Model (SM) is very successful in confronting the
data coming from the highest energy accelerators.
Still, there are theoretical reasons to expect that it is not complete, and
one of the first questions in the quest for new physics is
what is the relevant scale, where new phenomena
can give experimental signatures.
Recently, a radical proposal~\cite{ADD,ADD2,ADD3} has been
put forward for the solution of the hierarchy problem, which brings
close the electroweak scale $\rm m_{EW} \sim 1\; TeV$ and the
Planck scale $\rm M_{Pl} = \frac{1}{\sqrt{G_N}} \sim 10^{15}\; TeV$.
In this framework the effective four-dimensional $\rm M_{Pl}$ is
connected to a new $\rm M_{Pl(4+n)}$ scale in a (4+n) dimensional
theory:
\Be
\rm M_{Pl}^2 \sim M_{Pl(4+n)}^{2+n} R^n
\Ee
where there are n extra compact spatial dimensions of radius
$\rm \sim R$. This can explain the observed weakness of
gravity at large distances. At the same time,
quantum gravity becomes strong at a scale M of the order of 1~TeV
and could have observable signatures at present and future colliders.

The first experimental searches for large extra dimensions
have concentrated on the effects of real and virtual graviton
emission\footnote{For searches in Bhabha scattering
see e.g.~\cite{l3gr1,l3gr2,Bourilkov:1999}.}.
In a string theory of quantum gravity~\cite{Antoniadis,Peskin}
there are additional modifications of Standard Model amplitudes
and new phenomenological consequences.
Effective contact interactions caused by massive string mode
oscillations might compete with or even become stronger than those
due to virtual exchange of Kaluza-Klein excitations of gravitons,
and thus provide the first signature of low scale gravity or
a lower bound on the string scale.

Bhabha scattering above the Z resonance offers a reach hunting
field for new phenomena~\cite{Bourilkov:1998,Bourilkov:1999}.
It can be used to search for manifestations of contact
interactions and as a very sensitive probe of the point-like
structure of the electron.

This paper is organized as follows.
In sections 2 and 3 the experimental data and the analysis technique
are presented.
In the following section, we describe the search for
effects of TeV strings in Bhabha scattering.
In sections 5 and 6 we use the data to obtain limits on
the scale of different contact interaction models, and on
the size of electrons respectively.
We conclude with a discussion of the results.

%
%
\section*{Experimental Data}

Data on fermion-pair production at 183 or 189 GeV centre-of-mass energies
from the LEP2 collider has become available recently.
In the following we will concentrate on the measurements of
Bhabha scattering at these two
highest energy points, where large data samples have been
accumulated during the very successful LEP runs in 1997 and 1998.

The ALEPH~\cite{al183}, L3~\cite{l3189} and OPAL~\cite{op183,op189}
collaborations have presented results for the differential cross
section of Bhabha scattering.
In the case of L3 and OPAL the results are for both
energy points and the scattering angle $\rm \theta$ is
the angle between the incoming and the outgoing electrons
in the laboratory frame.
In the ALEPH case the measurements are at 183 GeV
and the scattering angle is defined
in the outgoing $\rm e^+e^-$ rest frame.
The acceptance is given by
the angular range $\rm |\cos\theta|<0.9$
for the ALEPH and OPAL measurements and by
$\rm  44^{\circ}<\theta<136^{\circ}$
for the L3 measurement.

The experiments use different strategies to isolate the
high energy sample, where the interactions take place at energies
close to the full available centre-of-mass energy. This
sample is the main search field for new physics.
L3 and OPAL apply an acollinearity cut of $\rm 25^{\circ}$
and $\rm 10^{\circ}$ respectively.
ALEPH defines the effective energy, $\rm \sqrt{s'}$, as the invariant
mass of the outgoing fermion pair. It is determined from the
angles of the outgoing fermions.
For details of the selection procedures, the statistical and
systematic errors we refer the reader to the publications of
the LEP experiments.

%
%
\section*{Analysis Method}

The Standard Model predictions for the differential cross sections of
Bhabha scattering at 183 and 189 GeV are computed with
the Monte Carlo generator BHWIDE~\cite{BHWIDE}.
We assign a theory uncertainty of 1.5~\% on the absolute scale
of the predictions.
In all cases the individual experimental
cuts of the selection procedures and the isolation of the high
energy samples are taken into account.
The results are cross-checked with the semi-analytic program
TOPAZ0~\cite{TOPAZ0}.

The effects of new phenomena are computed as a function of
a generic parameter $\rm \varepsilon$, defined for
each individual case in the corresponding section.
Initial-state radiation (ISR) changes the effective centre-of-mass
energy in a large fraction of the observed events.
We take these effects into account by computing the first order
exponentiated differential cross section following~\cite{Kleiss}.
Other QED and electroweak corrections give smaller effects and are neglected.

In total we have 47 data points: 28 from the 3 differential
spectra at 183 GeV and 19 from the L3 and OPAL spectra at
189 GeV.
A fitting procedure similar to the one
in~\cite{Bourilkov:1999,sneut} is applied.

A negative log-likelihood function is constructed by combining
all data points at the two centre-of-mass energies:
\Be
\rm -\log \mathcal{L} = \sum_{r=1}^{n}\left(\frac{(Prediction(SM, \varepsilon) - Measurement)^2}{2 \cdot \Delta_{Measurement}^2}\right)_r
\label{eqll1}
\Ee
\begin{eqnarray}
\rm \Delta_{Measurement}& = &\rm error(Prediction(SM, \varepsilon) - Measurement)
\label{eqll2}
\end{eqnarray}
where $ Prediction(SM, \varepsilon)$ is the SM expectation for a
given measurement (a point in the differential spectra) combined with 
the additional effect of new phenomena as a function of the mass scale
or electron size,
and $ Measurement$ is the corresponding measured quantity.
The index $\rm r$ runs over all data points.
The error on a deviation consists of three parts, which are combined in
quadrature: a statistical error and a systematic error (as given by the
experiments) and the theoretical error assigned above.
The systematic errors account for small correlations between
data points.

%
%
\section*{TeV Strings in Bhabha Scattering}

In~\cite{Peskin} the authors develop a model to study the effects
of string Regge excitations on physical cross sections by a
simple embedding  of the Quantum Electrodynamics of electrons and
photons into string theory. They use only one gauge group and only
vector-like couplings, in order to avoid complications but grasp
the general phenomenological picture. The results are model-dependent.

The effects of TeV scale strings on Bhabha scattering are computed
from the leading-order scattering amplitudes. All amplitudes are
multiplied by a common form-factor
\Be
\rm \mathcal{S}(s,t) = \frac{\Gamma(1-\frac{s}{M_S^2})\Gamma(1-\frac{t}{M_S^2})}{\Gamma(1-\frac{s}{M_S^2}-\frac{t}{M_S^2})}.
\Ee

In the case where the string scale $\rm M_S$ is close to or
smaller than the centre-of-mass energy, the Gamma-functions
in this form-factor produce a very reach and complicated
resonance structure.
On the other hand, in the limit where the Mandelstam variables
s and t are much smaller than $\rm M_S$, we have
\Be
\rm \mathcal{S}(s,t) = (1 - \frac{\pi^2}{6}\frac{st}{M_S^4} + ...).
\Ee
So in this model the leading corrections are proportional to
$\rm M_S^{-4}$, corresponding to an operator of dimension 8.

\begin{figure}[htbp]
  \begin{center}
  \resizebox{0.85\textwidth}{0.60\textheight}{
  \includegraphics*{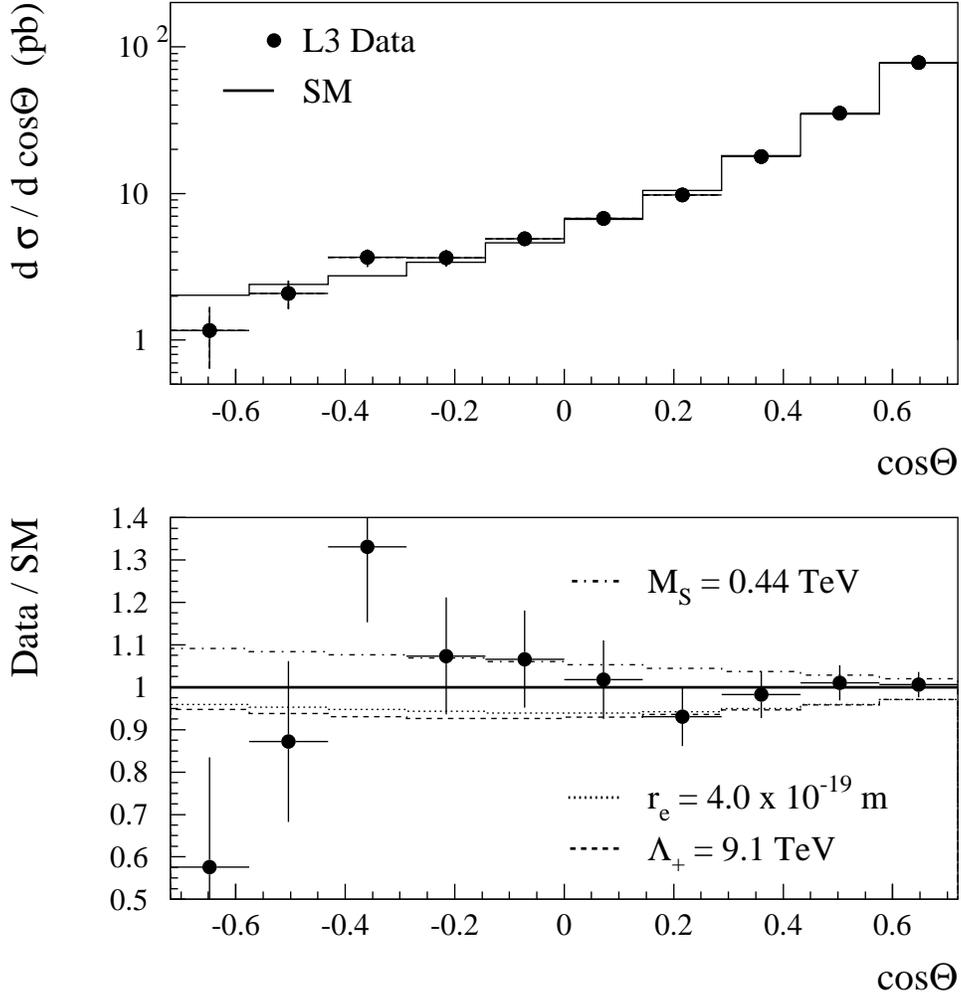}
  }
  \end{center}
  \caption{\em The differential cross section for Bhabha scattering
           as measured by the L3 collaboration at 189 GeV.
           The lower plot shows the ratio (data/SM expectation)
           together with the expected deviations from the SM for
           string models (dot-dash), finite electron size
           (dotted) and VV contact interactions (dashed).}
  \label{fig:figure1}
\end{figure}

To compare the string predictions to the data on Bhabha scattering
above the Z resonance one has to handle also the contributions due
to Z exchange and the interference with photon exchange amplitudes.
The Z is not part of the string QED model developed in~\cite{Peskin},
but as all QED Bhabha scattering amplitudes are multiplied by the
common factor $\rm \mathcal{S}(s,t)$, the authors suggest to
compare the differential cross section to the simple formula
\Be
\rm \frac{d \sigma}{d \COST} = (\frac{d \sigma}{d \COST})_{SM}\cdot |\mathcal{S}(s,t)|^2 .
\Ee

\begin{figure}[htbp]
  \begin{center}
  \resizebox{0.85\textwidth}{0.60\textheight}{
  \includegraphics*{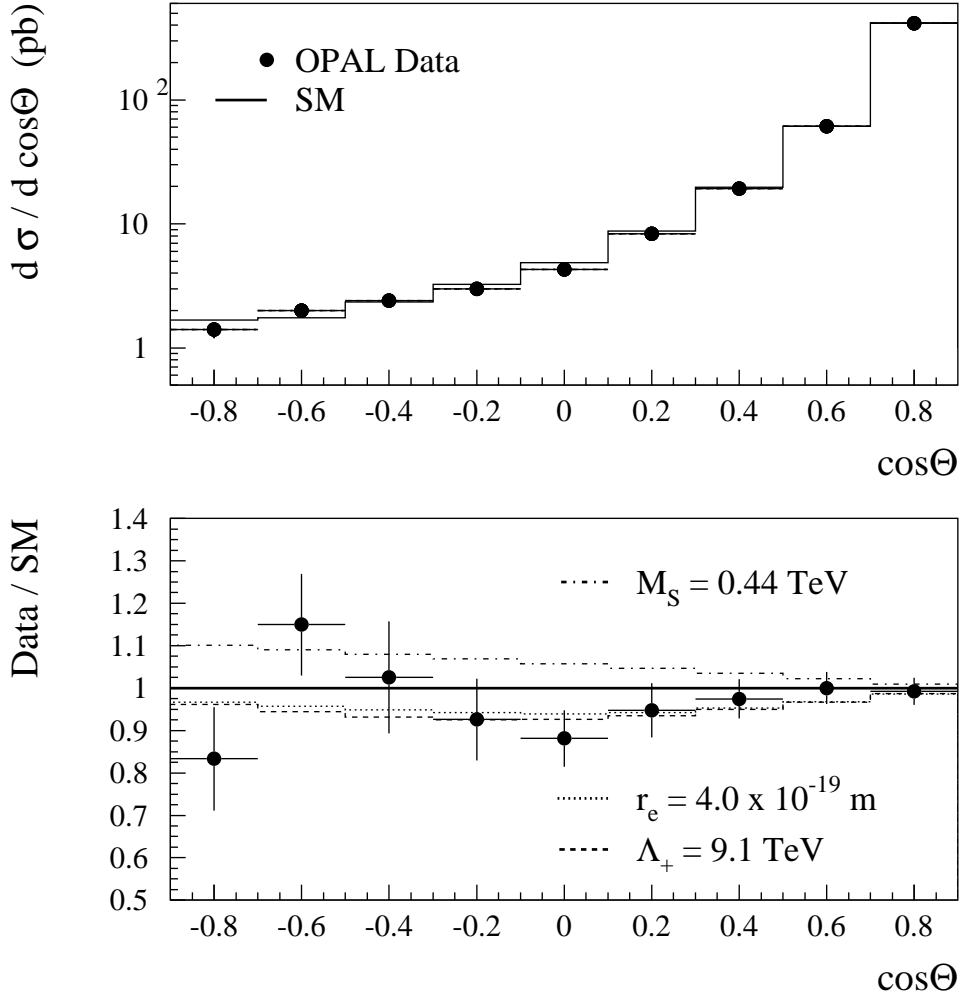}
  }
  \end{center}
  \caption{\em The differential cross section for Bhabha scattering
           as measured by the OPAL collaboration at 189 GeV.
           The lower plot shows the ratio (data/SM expectation)
           together with the expected deviations from the SM for
           string models (dot-dash), finite electron size
           (dotted) and VV contact interactions (dashed).}
  \label{fig:figure2}
\end{figure}

The data from the LEP collaborations at 183 and 189 GeV
show no statistically significant deviations from the SM
predictions due to string effects.
In their absence, we use the log-likelihood method, which after
proper normalization gives the confidence level for any value of
the scale $\rm M_S$ in the physically allowed region.
The exact definition can be found in~\cite{Bourilkov:1999}.
The one-sided lower limit on the scale $\rm M_S$ at 95\% 
confidence level is:
\Be
\rm M_S = 0.631\ TeV.
\Ee

Examples of the data analysis at 189 GeV are shown
in~\Figref{figure1} and~\Figref{figure2},
where the SM predictions and the expectations from several
manifestations of new phenomena are compared to the measurements
of the L3 and OPAL collaborations, respectively.
In these figures we plot the combined statistical and systematic
errors; the theory uncertainty is not shown.
In the area of the forward peak the theory uncertainty in the
SM prediction starts to limit the precision of our study.

%
%
\section*{Contact Interactions}

The standard framework, used in searches for deviations from the SM
predictions, is the most general combination of helicity conserving
dimension-6 operators~\cite{PeskinCI}.
In this scheme, new interactions beyond the Standard Model are
characterised by a coupling strength, $g$, and by an energy scale,
$\Lambda$, which can be viewed as the scale of compositeness. 
At energies much lower than $\Lambda$, we have an effective Lagrangian
leading to four-fermion contact interactions.

\begin{table}[h] 
 \renewcommand{\arraystretch}{1.2}
  \begin{center}
    \begin{tabular}{|c||c||cc|}
\hline
           &            & \multicolumn{2}{|c|}{$\EE$} \\   
~~Model~~  & Amplitudes &~~~~$\Lambda_-$~~~&~~~~$\Lambda_+$~~~ \\ 
           &$\rm [\eta_{LL}, \eta_{RR},\eta_{LR},\eta_{RL}]$
                        &~~~~[TeV]~~      &~~~~[TeV]~~~ \\
\hline  \hline
~~~~LL~~~~ & $\rm[\pm 1,     0,     0,     0]$ &  7.7 &  6.0  \\
RR         & $\rm[    0, \pm 1,     0,     0]$ &  7.6 &  6.0  \\
\hline                 
LR         & $\rm[    0,     0, \pm 1,     0]$ &  9.2 &  7.0  \\
RL         & $\rm[    0,     0,     0, \pm 1]$ &  9.2 &  7.0  \\
\hline                 
VV         & $\rm[\pm 1, \pm 1, \pm 1, \pm 1]$ & 16.2 & 13.0  \\
AA         & $\rm[\pm 1, \pm 1, \mp 1, \mp 1]$ &  8.0 & 10.4  \\
\hline                 
LL$+$RR    & $\rm[\pm 1, \pm 1,     0,     0]$ & 10.7 &  8.6  \\
LR$+$RL    & $\rm[    0,     0, \pm 1, \pm 1]$ & 12.9 & 10.1  \\
LL$-$RR    & $\rm[\pm 1, \mp 1,     0,     0]$ &  4.3 &  4.2  \\
\hline
\end{tabular}
  \end{center}
   \caption{
    Results of contact interaction fits to Bhabha scattering.
    The numbers in brackets are the values of 
    $\rm [\eta_{LL}, \eta_{RR},\eta_{LR},\eta_{RL}]$
    defining to which helicity amplitudes the contact interaction contributes.
    The models cover the interference
    of contact terms with single  as well as
    with a combination of helicity amplitudes.  
    The one--sided 95\% confidence level lower limits on the parameters
    $\Lambda_{+}$ ($\Lambda_{-}$) given in {\TeV}  
    correspond to the upper (lower) sign of 
    the parameters $\eta$, respectively.  
    }
  \label{tab:ci-leptons}
\end{table}

The differential cross section for fermion-pair production
in $\rm e^+e^-$ collisions can be decomposed in the usual
way as:
\Be
\rm \frac{d \sigma}{d \Omega} = SM(s,t)+\varepsilon\cdot C_{Int}(s,t)+\varepsilon^2\cdot C_{CI}(s,t)
\Ee
where $\rm SM(s,t)$ is the Standard Model contribution,
$\rm C_{CI}(s,t)$ comes from the contact interaction amplitude and
$\rm C_{Int}(s,t)$ is the interference between the SM and the 
contact interaction terms. The exact form of these functions is given
in~\cite{PeskinCI}.
By convention $\rm \frac{g^2}{4\pi}=1$ and
$\rm |\eta_{ij}|\leq 1$, where $\rm (i,j=\mathrm{L,R})$ labels
the helicity of the incoming and outgoing fermions.
We define 
\Be
\rm \varepsilon = \frac{g^2}{4\pi}\frac{sign(\eta)}{\Lambda^2}
\Ee
where the sign of $\rm \eta$ enables to study both the cases of positive
and negative interference.

\begin{figure}[htbp]
  \begin{center}
  \resizebox{0.80\textwidth}{0.50\textheight}{
  \includegraphics*{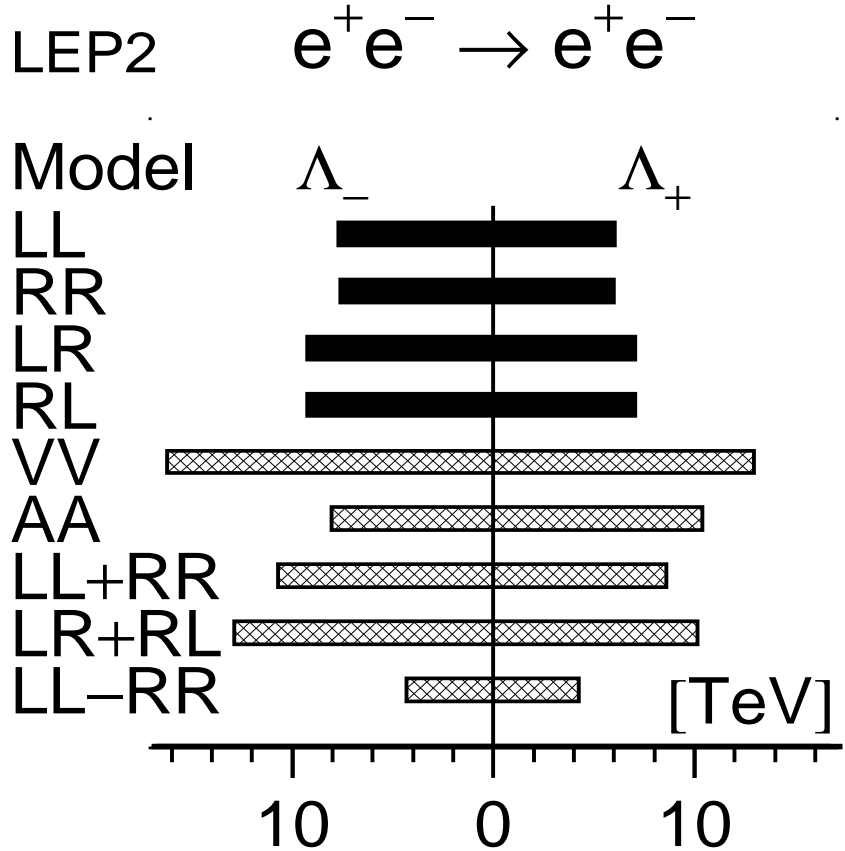}
  }
  \end{center}
  \caption{\em 
    One--sided 95\% confidence level lower limits on the scale $\Lambda_{+}$ 
    and $\Lambda_{-}$ for contact interactions in Bhabha scattering.}
  \label{fig:figure3}
\end{figure}

As discussed in the previous section,
the data from the LEP collaborations at 183 and 189 GeV
show no statistically significant deviations from the SM
predictions.
In their absence, using the same technique we derive one-sided
lower limits on the scale
$\rm \Lambda$ of contact interactions at 95\% confidence level.
They are summarized in Table~\ref{tab:ci-leptons} and~\Figref{figure3}.
The results presented here improve on the limits obtained by
individual LEP experiments~\cite{al183,l3ci,op189,de172}.

%
%
\section*{Electron Size}

In the Standard Model leptons, quarks and gauge bosons are
considered as point-like particles. A possible substructure
or new interactions at as yet unexplored very high energies
could manifest themselves as finite radii and anomalous
magnetic dipole moments of these particles.

The high precision measurements of the magnetic dipole
moment $\rm (g-2)_e$ of the electron can be used to put stringent
limits on the electron radius $\rm r_e$~\cite{Brodsky,Zerwas:1995}.
If non-standard contributions to
$\rm (g-2)_e$ scale linearly with the electron mass,
the bound is $\rm r_e \sim 2\cdot 10^{-23}\ m$.
On the other hand, if they scale quadratically with
the electron mass, which is a natural consequence of chiral
symmetry~\cite{Brodsky},
the bound is reduced to $\rm r_e \sim 3\cdot 10^{-18}\ m$.
In~\cite{Zerwas:1995} the authors perform an analysis
of the high precision data on the Z resonance, noting
that while the assumption of elementary photons is
quite natural, the same is less obvious for the very massive
Z bosons. In the pure electron case the limit is
not competitive with the $\rm (g-2)_e$ results.

Here we perform a new analysis based on the LEP2
data on Bhabha scattering, where again the photon
exchange gives the dominating amplitudes both in the
t- and s-channels,  and good
sensitivity to electron substructure can be expected.
The differential cross section for fermion-pair production
in $\rm e^+e^-$ collisions far above the Z is modified as:
\Be
\rm \frac{d \sigma}{d Q^2} = (\frac{d \sigma}{d Q^2})_{SM}\cdot F_e^2(Q^2)\cdot F_f^2(Q^2)
\Ee
where $\rm F_e$ and $\rm F_f$  are the form-factors of the initial (final)
state fermions. They are parametrized in the standard way
as~\cite{Zerwas:1995}:
\Be
\rm F(Q^2) = 1 + \frac{1}{6}\cdot Q^2\cdot r^2
\Ee
where $\rm Q^2$ is the Mandelstam variable s or t for
s- or t-channel exchange, and
$\rm r^2$ is the mean-square radius of the fermions.
This formalism is a convenient way to estimate the electron
size in the case where the product $\rm Q^2\cdot r^2$ is small.

From the data of the LEP collaborations at 183 and 189 GeV
we extract the following upper limit on the electron radius
at 95\% confidence level:
\Be
\rm r_e < 2.8\cdot 10^{-19}\ m.
\Ee

This limit is one order of magnitude lower than the limit
derived from $\rm (g-2)_e$ measurements in the case where
the deviations from the SM of the magnetic dipole moment of
the electron depend quadratically on its mass.

High energy analyses have been performed in interactions
involving electrons and quarks, assuming a single form-factor
for all fermions.
The H1 collaboration at HERA uses deep inelastic scattering
and obtains a limit of \mbox{$\rm r < 26\cdot 10^{-19}\ m$}
at 95~\% confidence level~\cite{h1-95}.
The CDF collaboration at the TEVATRON studies the Drell-Yan process
to put a limit of \mbox{$\rm r < 5.6\cdot 10^{-19}\ m$}
at 95~\% confidence level~\cite{cdf-97}.

%
%
\section*{Discussion}

The search for TeV strings motivates a fresh look at
Bhabha scattering. In the model analyzed here the
string realization of quantum gravity is manifested
as a form-factor which modifies the differential
cross section. The lower limit obtained in our analysis
of LEP2 data is \mbox{$\rm M_S = 0.631\ TeV$}.
In~\cite{Bourilkov:1999} from the study of 
virtual graviton exchange in gravity models with
large extra dimensions we obtained a lower limit on their 
scale of $\rm \Lambda_T = 1.412\ TeV$ for positive inteference
($\lambda = +1$)~\footnote{
This value of $\rm \Lambda_T$ corresponds, depending on
the convention, also to a gravity scale
$\rm M_s = 1.261\ TeV$. The gravity scale with subscript small s
should not be confused with the string scale $\rm M_S$,
studied here.}.
As noted in~\cite{Peskin}, the gravity scale is between
$\rm 1.6 \div 3.0 \cdot M_S$, depending on the coupling
strength. The results on the gravity scale from~\cite{Bourilkov:1999}
and on the string scale from this analysis agree well with
each other.

It is interesting
to note that our study of the electron size also
leads to form-factors modifying the differential cross
section, but with opposite sign. The limit derived
here, \mbox{$\rm r_e < 2.8\cdot 10^{-19}\ m$}, 
becomes \mbox{$\rm M_r > 0.705\ TeV$}, if translated
to a mass scale. This is a reflection of the similar magnitude
of the effects at LEP2 energies in both cases, even if the physics
mechanisms involved are different.

In the framework of contact interactions very stringent
bounds exceeding 10~TeV are obtained. When interpreting
the physical meaning of these limits, we should
remember that a strong coupling $\rm \frac{g^2}{4\pi}=1$
for the novel interactions is postulated by convention.
If we assume a coupling of electromagnetic strength,
the limits can be translated:
\Be
\rm \Lambda' = \sqrt{\alpha_{QED}}\cdot \Lambda = 0.085 \cdot \Lambda
\Ee
where we have used the value of the fine structure constant and
ignored the small effect of a running $\rm \alpha_{QED}$.
For instance the VV model with positive interference
gives effects similar to the ones resulting from a finite electron size,
as shown in~\Figref{figure1} and~\Figref{figure2}. 
The limit for the VV model translates as follows:
\Be
\rm \Lambda_{+} = 13.0\ TeV \Rightarrow \Lambda' = 1.1\ TeV \Rightarrow r = 1.8\cdot 10^{-19}\ m.
\Ee
This results is comparable with the upper limit for electron
substructure, derived using form-factors.

The measurements of Bhabha scattering above the Z resonance
confirm the predictions of the Standard Model  and
reach already a similar level of precision as the best
theoretical tools available.
In order to fully exploit the physics potential of the large data samples
collected during the LEP running in 1999 and expected in 2000,
improved theory predictions are very desirable.
Bhabha scattering is a probe, sensitive  enough to provide a first window
to new physics at the TeV scale.

%
%

%
%
\section*{Acknowledgements}
The author is grateful to A.~B\"ohm, M.~Peskin and I.~Antoniadis
for valuable discussions.

%
%

\bibliographystyle{/afs/cern.ch/l3/paper/biblio/l3stylem}
\bibliography{%
/afs/cern.ch/l3/paper/biblio/l3pubs,%
/afs/cern.ch/l3/paper/biblio/aleph,%
/afs/cern.ch/l3/paper/biblio/delphi,%
/afs/cern.ch/l3/paper/biblio/opal,%
/afs/cern.ch/l3/paper/biblio/markii,%
/afs/cern.ch/l3/paper/biblio/otherstuff,%
eth-pr-00-xx}

\end{document}